\documentclass[aps,prd,twocolumn,preprintnumbers,superscriptaddress,floatfix,amsmath,amssymb,showkeys]{revtex4}

\usepackage{multirow, graphicx,amssymb,url,mathrsfs,amsmath,amsfonts}
\usepackage{eucal,wrapfig,setspace}
\usepackage{amsxtra,amstext,latexsym}
\usepackage{pdflscape}
\usepackage{slashed}
\usepackage{graphicx}
\usepackage{xcolor}

\usepackage{caption}
\usepackage{sidecap}
\usepackage{dutchcal}
\usepackage{calligra}
 \newcommand{\scrn}{\mbox{${\mathscr N}$}}

\def\IR{{\hbox{{\rm I}\kern-.2em\hbox{\rm R}}}}
\def\IB{{\hbox{{\rm I}\kern-.2em\hbox{\rm B}}}}
\def\IN{{\hbox{{\rm I}\kern-.2em\hbox{\rm N}}}}
\def\IC{\,\,{\hbox{{\rm I}\kern-.59em\hbox{\bf C}}}}
\def\IZ{{\hbox{{\rm Z}\kern-.4em\hbox{\rm Z}}}}
\def\IP{{\hbox{{\rm I}\kern-.2em\hbox{\rm P}}}}
\def\IH{{\hbox{{\rm I}\kern-.4em\hbox{\rm H}}}}
\def\ID{{\hbox{{\rm I}\kern-.2em\hbox{\rm D}}}}

\newcommand{\beq}{\begin{equation}}
\newcommand{\eeq}{\end{equation}}
\newcommand{\bea}{\begin{eqnarray}}
\newcommand{\eea}{\end{eqnarray}}

\begin{document}
%%%%%%%%%%%%%%%%%%%%%%%
\renewcommand{\topfraction}{1.0}
\renewcommand{\bottomfraction}{1.0}
\renewcommand{\textfraction}{0.0}
%%%%%%%%%%%%%%%%%%%%%%%

\newcommand\sect[1]{\emph{#1}---}

\title{Any Room Left for Technicolor? Dilepton Searches at the LHC and Beyond}

\author{Alexander Belyaev}
\email{a.belyaev@soton.ac.uk}
\affiliation{STAG Research Centre and  Physics \& Astronomy, University of
Southampton, Southampton, SO17 1BJ, UK}
\affiliation{Particle Physics Department, Rutherford Appleton Laboratory, Chilton, Didcot, Oxon OX11 0QX, UK}

\author{Azaria Coupe}
\email{A.D.Coupe@soton.ac.uk}
\affiliation{STAG Research Centre and  Physics \& Astronomy, University of
Southampton, Southampton, SO17 1BJ, UK}

\author{Nick Evans}
\email{evans@soton.ac.uk} 
\affiliation{STAG Research Centre and  Physics \& Astronomy, University of
Southampton, Southampton, SO17 1BJ, UK}

\author{Daniel Locke}
\email{D.Locke@soton.ac.uk}
\affiliation{STAG Research Centre and  Physics \& Astronomy, University of
Southampton, Southampton, SO17 1BJ, UK}

\author{Marc Scott}
\email{ms17g08@soton.ac.uk}
\affiliation{STAG Research Centre and  Physics \& Astronomy, University of
Southampton, Southampton, SO17 1BJ, UK}


\begin{abstract}
Precision electroweak data, a light higgs and LHC searches for new spin one particles are all very constraining on technicolor models. We use a holographic model of  walking techicolor (WTC) gauge dynamics, tuned to produce a light higgs and low $S$ parameter, to estimate the range of possible vector($\rho$) and pseudo-vector($A$) resonance masses and couplings as a function of the number of colours and the number of flavours of techni-singlet and techni-doublet quarks. The resulting models predict techni-hadron masses and couplings  above the current 
limits from dilepton resonance searches at the LHC because their masses are enhanced by the strong coupling extending into the multi-TeV range, while couplings to Standard Model fermions are partly suppressed. The models emphasize the contortions needed to continue to realize technicolor, {the need to explore new signatures beyond dilepton for LHC and also motivate a 100 TeV proton collider.}
\end{abstract}

\maketitle

\section{Introduction}
Technicolor models of electroweak symmetry breaking \cite{Weinberg:1975gm} solve the hierarchy problem 
{by naturally generating a strong coupling regime in the TeV energy range and a resulting composite higgs}. 
They predict a large bound state spectrum. LHC data {has been used to study dilepton constraints on the parameter space  of technicolor vector and axial vector resonances in \cite{Belyaev:2008yj}, recently updated in \cite{Belyaev:2018qye}. To date that analysis have been presented in a large phenomenological parameter space. Our goal here is to study where UV complete models are likely to lie in that parameter space, whether the constraints are tough enough to exclude the paradigm, and to motivate further analysis or colliders that might do so. }

Strongly coupled models have been pressured by the precision electroweak data~\cite{Peskin:1990zt} (which warns against extended electroweak sectors) and the discovery of a light higgs \cite{Chatrchyan:2012xdj}. However, given our paucity of tools for computing in strongly coupled environments, there has been a hope that within the space of strongly coupled gauge theories are some that might still be tuned to the data. Walking theories \cite{Holdom:1981rm}, that lie close to the edge of the conformal window in gauge theories with varying $N_c$ and $N_f$, represent a sensible argument that such fine tuned models may exist with both small higgs mass \cite{Appelquist:1998xf} and electroweak precision data $S$ parameter \cite{Sundrum:1991rf}. 

A clear prediction of these models though is their large bound state spectrum which must emerge close to the electroweak scale. Light pseudo-Goldstone modes could, but need not, exist and when they do are hard to pin down because their mass is determined by breaking of the chiral symmetries by the potentially unknown origin of flavour physics (which could be strong). We will therefore concentrate on the vector($\rho$) and pseudo-vector($A$) mesons of such theories which are probably more robust in their mass predictions.

As a test case, we will present our theoretical predictions for SU($N_c$) technicolor theories with $N_f$ flavours in the fundamental representation. 
There is a choice as to how many of these $N_f$ flavours form SU(2)$_L$ doublets. The $S$ parameter suggests more than one doublet is unlikely. Further if more than one doublet contributes to determining the electroweak scale through $F_\Pi$ then the entire scale of the technicolor theory moves down potentially making these states more accessible. If there is a single doublet then there will be just a single triplet of spin one particles (each of the $\rho$ and $A$) that are most easily experimentally accessible (since they mix with the W and Z and can be produced singly in electroweak processes). Placing constraints on the single doublet theory therefore offers techicolor the maximal chance of escape and we will concentrate on this (since such models will be the last ones standing). We will though also present results for the spectrum of theories with multiple electroweak doublets. 

We wish to predict the masses and decay constants of the $\rho$ and $A$ states in the space of strongly coupled models. Lattice techniques have begun the job \cite{DeGrand:2015zxa} but they are computationally hard when the dynamics is spread over a wide range of scales and it will take many years of hard work to understand the full $N_c,N_f$ parameter space. To make progress more quickly we will describe the dynamics using holography \cite{Maldacena:1997re}. Holography provides a rigorous method of computation in a selection of strongly coupled gauge theories close to {$\scrn=4$ supersymmetric} gauge theory including theories with quarks \cite{Karch:2002sh}. In the quenched (probe) limit the key ingredient to determine the spectrum is the running anomalous dimension of the quark bilinear ($\bar{q}q$), $\gamma$ \cite{Jarvinen:2011qe}. Embracing that observation we can construct holographic models of generic gauge theories \cite{Alho:2013dka}. The predictions for the QCD ($N_c=N_f=3$) spectrum lie surprisingly close to observation at the 10$\%$ level and one can hope as one moves away to theories with e.g. walking behaviour the models will continue to make sensible predictions of the spectrum. For the purist  the approach we use lays down a ball park estimate and challenges them to estimate the parameter space of the models better. 

For generic $N_c,N_f$ the spectrum will look QCD-like with a heavy $\sigma$ (higgs) and a large value for the $S$ parameter. The ``last hope'' for technicolor (which one might hope to exclude) is that the (unknown) IR running is sufficiently fine tuned that it can generate a light higgs and low $S$.  The holographic models allow the $\sigma$  to become light if the running around the chiral symmetry breaking scale is near conformal \cite{Alho:2013dka}.  Most likely, if any, only a single choice of $N_c$ and $N_f$ will generate suitable walking and hence a suitably light composite higgs. Since we can't guess that theory we will instead allow every choice of theory (ie $N_f,N_c$) to have its best hope by tuning the IR running to generate a 125 GeV state. There is also a 5d gauge coupling in the holographic model that allows the $\rho$ and $A$ masses to be tuned together to achieve low values of $S$ ({this is the only way to achieve small $S$ in the simple holographic model presented}) - again we do this for all theories. Of course this means that the spectrum in most (if not all) cases will be wrong but our philosophy is to show where the theories might lie if treated most favourably to set the benchmark for total exclusion. 

To compare the predictions to data we will use the constraints placed on the phenomenological model of techni--$\rho$ and $A$ proposed in ~\cite{Foadi:2007ue}. The philosophy, based on the ideas of hidden local symmetry ~\cite{Bando:1984ej}, is to describe the vector mesons as the massive gauge bosons of a broken gauged SU($N_f$)$_L \otimes$SU($N_f$)$_R$ symmetry. The two main signals relevant for phenomenology
are Drell-Yan production and Vector Boson Fusion \cite{Belyaev:2008yj}. In each case a single $\rho$ or $A$ is produced through mixing with the electroweak gauge bosons via the combined mass matrix determined from the action. Constraints on this model (for the case of a single electroweak doublet), from Drell-Yan processes, have recently been updated in \cite{Belyaev:2018qye} to the March 2018 LHC results on CMS dilepton resonance search~\cite{Sirunyan:2018exx}. The holographic model makes predictions for the parameters of the phenomenological model so the constraints can be directly applied.

The results of the analysis in brief are that the technicolor theories that emerge are rather odd - they enter the strong coupling at a scale of 700 TeV or above before settling on an IR fixed point that trigger symmetry breaking at the 1 TeV or so range. We find the IR theory constructed in the way described is largely independent of the UV theory. The result is that the bound states of the theory know, through the strong interactions, of rather high scales and their holographic wave functions stretch to large UV scales. The result is that the $\rho$ and $A$ masses increase to $M_\rho \simeq 4$~TeV. Such theories,
with the specific couplings we have found, are beyond the reach of the current LHC dilepton searches. However, they do motivate new signatures to explore and future colliders with higher energies which could probe these scales.  In a sense such theories display the issues that any extension of the standard model that addresses the hierarchy problem must now encounter - to make the higgs light there must be tuning at one part in 100 or so and new states must be pushed to high scale. 

In the next two sections we will review our holographic model used to estimate the parameter space for technicolor models and the phenomenological analysis of~\cite{Belyaev:2018qye} before bringing the two together to show the exclusion in section 3. The act of forcing a small $S$ parameter in the holographic model corresponds to enforcing {$\rho$-$A$} degeneracy and this places the models in the parameter space of the phenomenological model where a measure of that degeneracy, $\mathcal{a}$, is close to zero (that the holographic predictions match the model's parameter space is tested by this fact). Unfortunately this is the toughest edge of the parameter space to probe experimentally. The reader who wishes to cut to the chase should {inspect Figure~\ref{fig:collider} where the bounds on the models in the coupling versus $A$ mass parameter space are shown with the} { holography} {predictions for the parameter space of technicolor overlayed. }

\section{Holographic Model}

Our holographic model is the Dynamic AdS/QCD model which is described in detail in \cite{Alho:2013dka}. The action is
\begin{equation} \begin{array}{lcl}
S & = & -\int d^4x~ d u, {\rm{Tr}}\, u^3 
\left[  {1 \over r^2} |D X|^2 \right.  \\ &&\\
&& \left. \hspace{1cm}  +  {\Delta m^2(r) \over u^2} |X|^2   + {1 \over 2 \kappa^2} (F_V^2 + F_A^2) \right], 
\label{daq} \end{array}
\end{equation}
Here {$u$ is the holographic coordinate dual to energy scale,} and $X$ is a field dual to the quark condensate $\bar{q} q$. The solution of its equation of motion, which can be found numerically, describes the vacuum of the theory. We pick the on mass shell condition $|X|(u=X_0) = X_0$ with $|X|'(X_0)=0$ and require $|X|=0$ in the UV so all techniquarks are massless. Fluctuations of $X$ describe the $\sigma$ and $\pi$ fields. 

The vector and axial vector fields describe the operators $\bar{q} \gamma^\mu q$ and $\bar{q} \gamma^\mu \gamma_5 q$ and their fluctuations give the $\rho$ and {$A$} spectrum and couplings.  

The theory lives in a geometry
\begin{equation} ds^2 = r^2 dx_{3+1}^2 + {1 \over r^2} du^2, ~~~~~~~r^2 = u^2 + |Tr X|^2 \end{equation} 
 $|Tr X|$ is included in the definition of $r$ in the metric which provides a ``back-reaction'' on the metric in the spirit of probe brane models \cite{Karch:2002sh} and communicates the mass gap to the mesonic spectrum. 

$\Delta m^2$ is a renormalization group scale/radially dependent mass term which can be fixed, for example, from the two loop running of the gauge coupling in the theory of interest as described in \cite{Alho:2013dka} - this ansatz includes IR fixed points for the running for appropriate choices of $N_c,N_f$. 
 
The spectrum of the theory is found by looking at linearized fluctuations of the fields about the vacuum where fields generically take the 
form $f(u)e^{ip.x}, p^2=-M^2$. A Sturm-Louville equation results for $f(u)$ leading to a discrete spectrum. By substituting the wave functions back into the action and integrating over $u$ the decay constants can also be determined.
The normalizations of the fluctuations are determined by matching to the gauge theory expectations for the vector-vector, axial-axial and scalar-scalar correlators in the UV of the theory. This full procedure is described in detail in \cite{Alho:2013dka}. Note that in the holographic literature \cite{{Alho:2013dka}} the dimension 2 coupling between the vector meson and it's associated source is normally written as $F_V^2$ whilst in the Weinberg sum rule literature \cite{Peskin:1990zt} it is written as $m_V F_V$. We will adopt the latter definition here to fit the other literature on technicolor.

Our models will focus first on a single electroweak doublet of techni-quarks but we will assume the existence of technicolor singlet quarks to change the UV running of the coupling. In the computations of $f_\pi$ and $F_{V/A}$ for the electroweak physics only the electroweak doublet contributes - so factors of $N_f$ and $N_c$ in these quantities reflect the values in a one doublet model.  As discussed in the introduction we will further tune the IR running of $\gamma$ in all our theories to generate a 125~GeV $\sigma$ meson.
To achieve this we set a value of $\alpha_{TC}$ where we deviate from the
UV running. Below that scale we allow $N_f$ to become a
free parameter and pick a $N_f^{IR}$ (which we stress is not
the true value of $N_f$ in the theory - in practice it is very similar for all cases and lies at 11.43) to let us tune the
$\sigma$ mass to the observed higgs mass value. 
\begin{figure}[htb]
   {\includegraphics[width=0.5\textwidth]{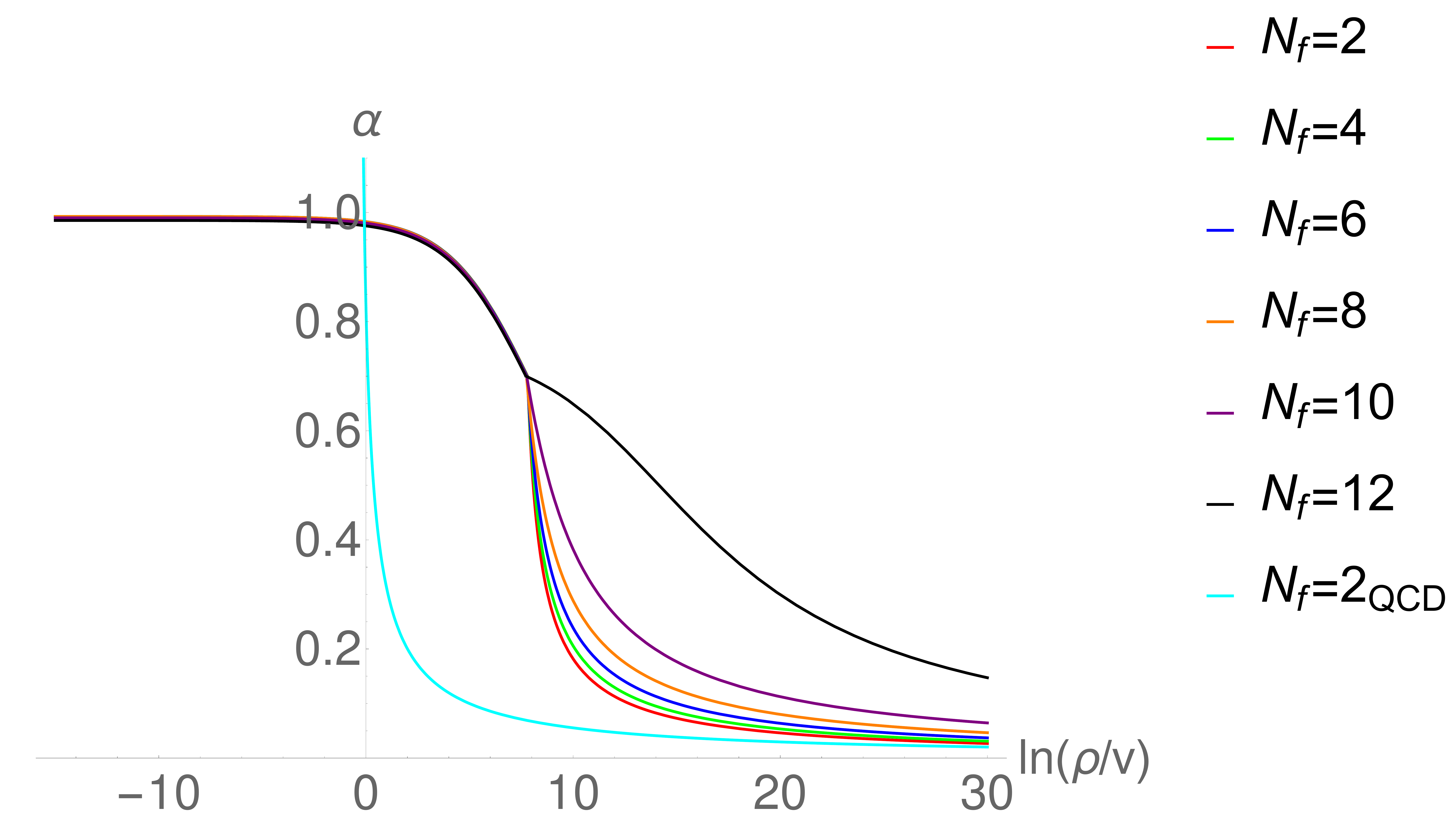} } 
  \caption{\label{fig:running}The running of $\alpha_{TC}$ against RG scale imposed on the holographic model with $N_c=3$. The curve furthest to the left is for a technicolor model that is a scaled up version of QCD with the usual two loop result for the running. The next curve over is that same theory forced to have a IR fixed point to produce a light higgs (clearly we know for this theory that this assumption is wrong!). Moving further to the right we see the running as further singlet techi-quarks are added, again with $N_f^{IR}$ chosen to give a light higgs. The IR of all such theories is shared and uniquely determined by needing the observed higgs mass.
  }
\end{figure}  
This
matching scale becomes a discontinuity in the running of $\alpha_{TC}, \gamma$
or the AdS scalar mass. In practice we deal with this
by performing all computations in sections and matching
the value of fields and their derivatives at the boundary
point. To provide an estimate of the errors on the spectrum
we allow the matching point in $\alpha_{TC}$ to vary from 0.3
to 0.7. In Figure~\ref{fig:running} we show an example of the running in the theories we impose - clearly they all share essentially the same IR
which is fixed by the higgs mass value. 
We will discuss the implications further in the final section. 
In the same spirit we will tune the coupling $\kappa$ in the model to produce {$\rho$--$A$} degeneracy to ensure the electroweak $S$ parameter 
\begin{equation}
S=4\pi\left[\frac{F_V^2}{M_V^2}-\frac{F_A^2}{M_A^2}\right] \ , \label{eq:WSR0} 
\end{equation}
is sufficiently small  (we pick $S=0.1$ as a benchmark point), even though this will not actually be the case for most $N_c,N_f$ theories. We are leaning over backwards to keep technicolor alive of course, but to understand a total exclusion on the parameter space this is sensible. Equally the models  display the large tunings needed for viability. Note tuning $\kappa$ to zero makes the Lagrangian terms for the $\rho$ and $A$ the same so that the $A$ mass drops to that of the $\rho$. However, since the suppressed, first term in the action is the one which links the symmetry breaking $X$ to the $A$, to maintain $f_\pi^2$ (which is the leading value in the $AA$ correlator) one must raise the overall scale. This is the main source of the rise in the masses relative to a QCD-scaled up theory. 

The parameter count in the holographic model is: for a particular theory with $N_c,N_f$ the UV running of $\alpha$ (and hence the anomalous dimension $\gamma$) is fixed by the perturbative two loop result. The overall scale is set by requiring $F_\Pi=246$~GeV.  We then modify the IR running - we change it at scales below some matching value of $\alpha_{TC}^{\rm match}$ (which we vary from 0.3 to 0.7 to provide the range of predictions, displayed as the {horizontal width of the prediction curves in Figure~\ref{fig:collider}) } by adjusting the effective value of $N_f$ in the IR and adjusting it to fix the $\sigma$ meson mass to the observed higgs mass. The model then predicts $M_\rho, F_\rho, M_A, F_A$ as a function of the 5d gauge coupling, $\kappa$. We tune $\kappa$ to give $S=0.1$. The remaining three predictions we will express as
\begin{equation} M_A,  ~~~~~ \tilde{g} = {\sqrt{2} M_V \over F_V},   ~~~~  \omega = {1 \over 2} \left({F_\pi^2 + F_A^2 \over F_V^2 } - 1  \right). \end{equation}
In fact for all our models $\omega < 0.05$ which is at a level where the experimental constraints are unchanged in  the high energy reach regime so we suppress that parameter in our plots.

\section{Phenomenological Model} 
The phenomenological model of the spin {one} states made from the electroweak doublet is \cite{Foadi:2007ue}
\begin{eqnarray}
{\cal L}_{\rm boson}&=&-\frac{1}{2}{\rm Tr}\left[\widetilde{W}_{\mu\nu}\widetilde{W}^{\mu\nu}\right]
-\frac{1}{4}\widetilde{B}_{\mu\nu}\widetilde{B}^{\mu\nu}\nonumber \\ &
-&\frac{1}{2}{\rm Tr}\left[F_{{\rm L}\mu\nu} F_{\rm L}^{\mu\nu}+F_{{\rm R}\mu\nu} F_{\rm R}^{\mu\nu}\right] \nonumber \\
&+& m^2\ {\rm Tr}\left[C_{{\rm L}\mu}^2+C_{{\rm R}\mu}^2\right]
+\frac{1}{2}{\rm Tr}\left[D_\mu M D^\mu M^\dagger\right] \nonumber \\ &
-& \tilde{g^2}\ r_2\ {\rm Tr}\left[C_{{\rm L}\mu} M C_{\rm R}^\mu M^\dagger\right] \nonumber \\
&-&\frac{i\ \tilde{g}\ r_3}{4}{\rm Tr}\left[C_{{\rm L}\mu}\left(M D^\mu M^\dagger-D^\mu M M^\dagger\right) \right.\nonumber \\ 
&+&  \left. C_{{\rm R}\mu}\left(M^\dagger D^\mu M-D^\mu M^\dagger M\right) \right] \nonumber \\
&+&\frac{\tilde{g}^2 s}{4} {\rm Tr}\left[C_{{\rm L}\mu}^2+C_{{\rm R}\mu}^2\right] {\rm Tr}\left[M M^\dagger\right]\nonumber \\ &
+& \frac{\mu^2}{2} {\rm Tr}\left[M M^\dagger\right]-\frac{\lambda}{4}{\rm Tr}\left[M M^\dagger\right]^2
\label{eq:boson}
\end{eqnarray}
where $\widetilde{W}_{\mu\nu}$ and $\widetilde{B}_{\mu\nu}$ are the ordinary electroweak field strength tensors, $F_{{\rm L/R}\mu\nu}$ are the field strength tensors associated to the vector meson fields $A_{\rm L/R\mu}$~\footnote{In Ref.~\cite{Foadi:2007ue}, where the chiral symmetry is SU(4) there is an additional term whose coefficient is labelled $r_1$. With an SU($N$)$\times$SU($N$) chiral symmetry this term is just identical to the $s$ term.}, and the $C_{{\rm L}\mu}$ and $C_{{\rm R}\mu}$ fields are
$C_{{\rm L}\mu}\equiv A_{{\rm L}\mu}-\frac{g}{\tilde{g}}\widetilde{W_\mu}$ and \\ $
C_{{\rm R}\mu}\equiv A_{{\rm R}\mu}-\frac{g^\prime}{\tilde{g}}\widetilde{B_\mu}
$

The matrix $M$  takes the form
\begin{eqnarray}
M=\frac{1}{\sqrt{2}}\left[v+H+2\ i\ \pi^a\ \tau^a\right]\ ,\quad\quad  a=1,2,3
\end{eqnarray}
Here $\pi^a$ are the Goldstone bosons produced in the chiral symmetry breaking, $v=\mu/\sqrt{\lambda}$ is the corresponding VEV, and $H$ is the composite higgs. We assume the higgs has Standard Model yukawa couplings to the fermions. The covariant derivative is
\begin{eqnarray}
D_\mu M&=&\partial_\mu M -i\ g\ \widetilde{W}_\mu^a\ \tau^a M + i\ g^\prime \ M\ \widetilde{B}_\mu\ \tau^3\ . 
\end{eqnarray}
When $M$ acquires its VEV, the Lagrangian of Eq.~(\ref{eq:boson}) contains mixing matrices for the spin one fields. The mass eigenstates are the ordinary SM bosons, and two triplets of heavy mesons:$\rho$ and  $A$.

Including all the interactions with the electroweak gauge and higgs fields of dimension 4 needs six parameters: the mass, $m$ and coupling $\tilde{g}$ of the new gauge fields, the higgs VEV $v$, and three couplings $r_2, r_3$ and $s$. 
The model then predicts 
\begin{equation}
M_V^2 = m^2 + \frac{\tilde{g}^2\ (s-r_2)\ v^2}{4}, ~~~~~
M_A^2 = m^2 + \frac{\tilde{g}^2\ (s+r_2)\ v^2}{4} 
\label{eq:masses}
\end{equation}
and
\begin{equation}
F_V  =  \frac{\sqrt{2}M_V}{\tilde{g}} \ , 
F_A  =  \frac{\sqrt{2}M_A}{\tilde{g}}\chi \ ,
F_\pi^2  =  \left(1+2\omega\right)F_V^2-F_A^2 \ ,
\label{eq:FVFAFP}
\end{equation}
where
\begin{eqnarray}
\omega \equiv \frac{v^2 \tilde{g}^2}{4 M_V^2}(1+r_2-r_3) \ , \quad \quad 
\chi \equiv 1-\frac{v^2\ \tilde{g}^2\ r_3}{4 M_A^2} \ . \label{eq:chi}
\end{eqnarray}
{Without loss of generality we chose $s=0$ here, noting that:
a) the $Z'/Z''$ 
production rates, as well as the partial decay width of
Z to fermions (di-jets and di-leptons) are independent of $s$ (at the per-mil level);
b)  the branchings of $Z'$ to} {dileptons} {increases by 10$\%$ at most for
$s$ reaching 10 in absolute value because of the  $Z'\to ZH$  partial width decreases;
c) we do not involve here higgs boson phenomenology and use only the dilepton channel to probe the WTC space.
%Setting $s=0$ therefore still provides robust constraints on the theory.
}

Of the five remaining variables we set $F_\Pi=$246~GeV, and $S=$0.1. This leaves three degrees of freedom $M_A, \tilde{g}, \omega$ which can be experimentally constrained.

{We have implemented the model in CalcHEP~\cite{Belyaev:2012qa} using LanHEP~\cite{Semenov:2014rea} to derive the Feynman rules~\cite{Belyaev:2008yj,Belyaev:2018qye}. The implementation of the model is publicly available at HEPMDB database~\cite{hepmdb} under hepmdb:1012.0102 ID.
In this implementation
we have extended previous implementation~\cite{Belyaev:2008yj}(hepmdb:1012.0102) by nonzero $s$ and $\omega$ parameters to be interpreted in the context of  the holographic description.
}

The two main signals relevant for phenomenology were shown to be Drell-Yan production and Vector Boson Fusion. In each case a single $\rho$ or $A$ is produced through mixing with the electroweak gauge bosons via the combined mass matrix determined from the action. {The Drell-Yan analysis }has recently been updated to the latest 13~TeV LHC data in \cite{Belyaev:2018qye}.

Phenomenologically the three parameters are treated as completely free parameters. 
The parameter count is the same as that of the holographic model which makes absolute predictions for these numbers as a function of $N_c, N_f$. We can therefore immediately superimpose the holographic predictions on the constraints from \cite{Belyaev:2018qye}.

We have explored the dependence of the experimental constraints on the parameter $\omega$. For $|\omega| < 0.3$ the impact on the exclusion regime is small and any changes occur at $M_A \simeq 1.5$~TeV. The high mass reach area is least {affected}. Given the holographic models place $\omega < 0.05$ in all cases we will simply suppress this parameter which is not playing a significant role in constraining the models.

A further useful parameter to monitor (although it is not independent) is  $\mathcal{a}$ from
\begin{equation} \mathcal{a} 4 \pi^2 F_\pi^4 = F_\rho^2 M_\rho^2 - F_A^2 M_A^2 \end{equation}
which provides a monitor of the second Weinberg sum rule or equally the degeneracy of the $\rho-A$ pair. Since the holographic model ensures a small $S$ parameter precisely by such degeneracy it is not surprising the models lie near the $\mathcal{a}=0$ curve in the $\tilde{g}-M_A$ plane. Unfortunately this is the extreme of the parameter space analyzed in the phenomenological model previously which is hardest to probe experimentally. It is instructive to know it might be favoured in real models.

\section{Results}

Our first goal is to place the recent experimental limits on WTC \cite{Belyaev:2018qye} in the context of predictions for real models. 
In this paper we use LHC limits on dilepton resonance  searches only
and reinterpret them for WTC parameter space. The choice of dilepton signature is very well motivated since this is probably the most clean signature for search of the vector resonances. However, as we will see below,  it becomes less efficient in the region of large values of $\tilde g$ where couplings of  resonances to SM fermions are suppressed.

In Figure~\ref{fig:collider} we present the up-to-date LHC reach for the phenomenological WTC model, so the reader can see the current LHC potential to probe the model parameter space.
We use here the CMS { DY limits on $Z^{\prime}$ production at 13~TeV ($36 fb^{-1}$) from the dilepton (combined dielectron and dimuon) final state~\cite{Sirunyan:2018exx} for the reinterpretation to limits on the WTC parameter space. The CMS limit is expressed 
as a ratio, $R_{\sigma}$ = $\sigma(pp\rightarrow Z^{\prime}\rightarrow \ell^{+}\ell^{-})/\sigma(pp\rightarrow Z\rightarrow \ell^{+}\ell^{-})$, of $Z^{\prime}$ signal cross section in the dilepton final state to the cross section of a Z boson to the dilepton final state. CMS calculate this Z boson cross section to NNLO in the control region of 60$\leq m_{\ell^{+}\ell^{-}}$120~GeV. CMS present the limit as $R_{\sigma}$ to remove the dependency on the theoretical prediction of $\sigma(pp\rightarrow Z\rightarrow \ell^{+}\ell^{-})$ and correlated experimental uncertainties.} {Using this limit we have found 95$\%$ CL limits on the WTC $\tilde g-M_A$ parameter space 
for the $\rho$ and $A$ separately and then overlay them to find overall combination.
We have used CalcHEP to evaluate signal at tree-level and modified ZWPROD program~\cite{Accomando:2010fz} to evaluate mass-dependent QCD NNLO  K-factor. 
The current LHC observed  limit is indicated
by the combined shaded area in  Figure~\ref{fig:collider}.} {
One can clearly see that the LHC is currently not sensitive to the parameter space of WTC models predicted by holography, even those models with a large number of techni-doublets where $\tilde g\simeq 2.5$ and $M_A\simeq 4$~TeV. 

\begin{figure}[htb]
   \includegraphics[width=0.5\textwidth]{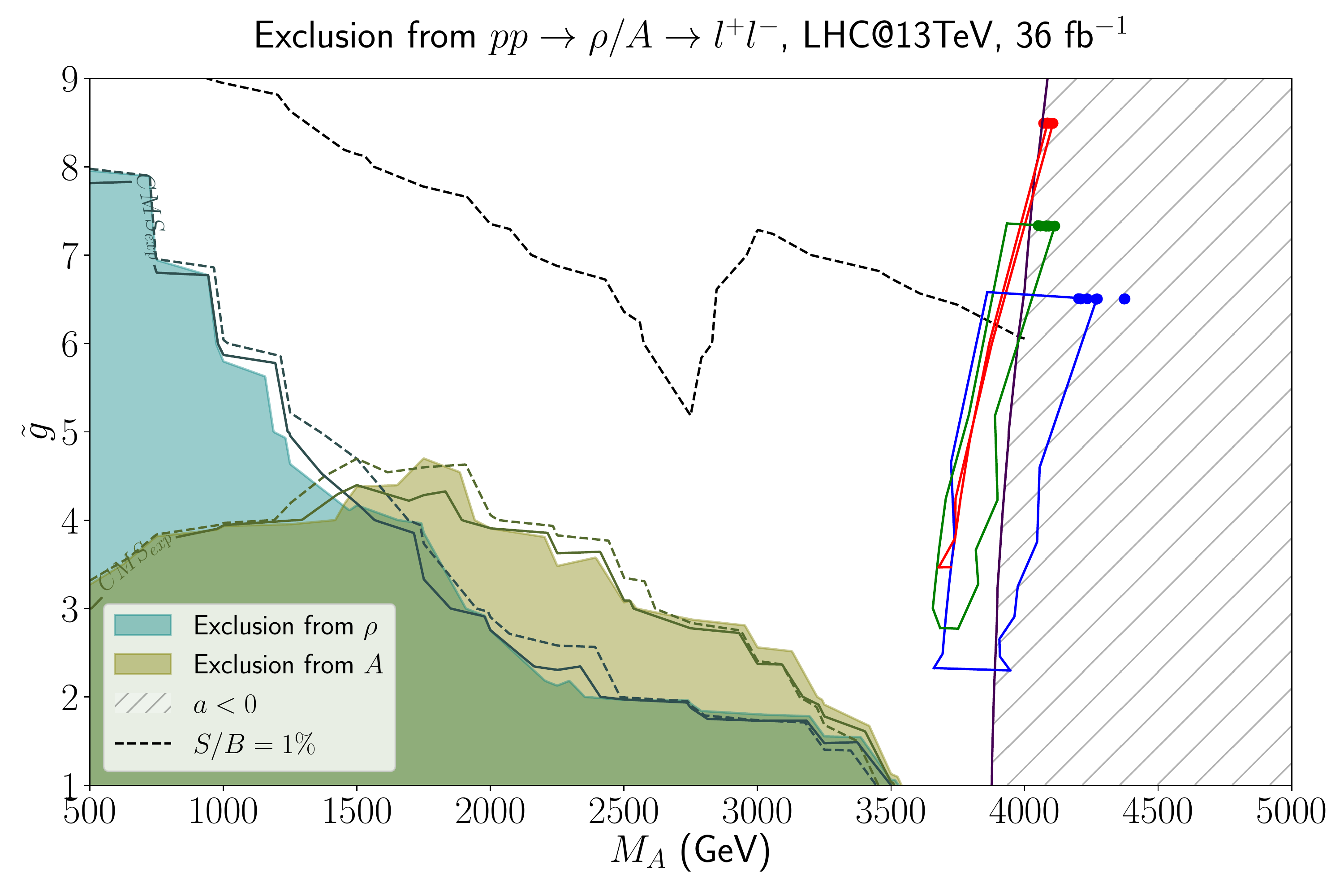}
  \caption{\label{fig:collider}
 Shaded areas present 95\% CL exclusion  on the $M_A- \tilde{g}$ plane 
 from the CMS observed  limit on  dilepton resonance searches at the LHC@13TeV with 36 fb$^{-1}$. Solid and dashed lines along the borders of the shaded area represent an expected CMS limit and our 
 limit  using binned likelihood method respectively.
 The predictions of our holographic model (tuned at each $N_c,N_f$ to give $S$=0.1 and the correct higgs mass) are overlaid.  The  red colour indicates $N_c=3$, green --- $N_c=4$ and blue --- $N_c=5$. The top edge of the box in each case is the one electroweak doublet theory result with the width representing an estimate of the theoretical error (we match the IR running at different values as described in the text). 
  The points correspond to the motion of the right hand point on that line as the number of singlets is changed to vary the UV running - the effect is small because the theories share much the same IR running to generate $m_h$. Moving down in the box corresponds to increasing the number of electroweak techni-doublets from one to $2 N_c$ where the theories are assumed to enter the conformal window. Parameter $\mathcal{a}$, from the phenomenological model, is related to $\rho-A$ degeneracy and the holographic points lie near the line $\mathcal{a}=0$ as a result of tuning to a small $S$ parameter. 
  } \end{figure}

Besides finding the observed limit as a reinterpretation of the CMS results we have closely reproduced an expected CMS limit (the solid lines along the borders of the shaded area
in Figure~\ref{fig:collider} to be compared with the respective dashed lines from our approach)
from the dilepton search in order to validate our approach and extend its use to projections for future collider energies and luminosities.

{We have evaluated our expected limits using a binned likelihood method. 
We assume resonance widths are negligible compared to the gaussian-smearing effects of finite detector resolution.} {The signal hypothesis pdf is defined by a Gaussian of width equal to the detector resolution (1.2\% of resonance mass), and a signal-strength modifier, $\mu$, which is the expected number of events at the experiment. Background is estimated by generating very high statistics for invariant dilepton mass distributions. Where there are few background events (e.g. $m_{\ell^{+}\ell^{-}}\geq$ 2~TeV at 13~TeV), we use the CL$_s$ method alongside a toy Mont\'e Carlo in order to construct the distribution of a single test-statistic for background only and signal+background hypotheses.

In Figure~\ref{fig:collider} we also present a dashed black line lying in the large $\tilde g$ region and indicating a 1\% level of signal-to-background ratio (from the most optimistic expected systematic uncertainty) as an indication of the absolute limit of the dilepton signature potential to probe the WTC paradigm.}
{This contour line is not expected to change with the increase of the collider energy since the irreducible dilepton background and the signal will scale the same way with the energy increase.}

These results have reach to 3.5 TeV in mass and couplings $\tilde{g} \sim 8$ so at first glance appear very constraining. However, let us first orient ourselves in theory space. QCD is a gauge theory that we are fully confident of the spectrum - we can therefore consider a techicolour model with an SU(3) gauge group and $N_f=2$ light quarks (up to the influence of the strange quark) by scaling up QCD. We scale $f_\pi=93$~MeV to $F_\Pi = 246$~GeV and find $M_\rho= 2.05$~TeV, $M_A= 3.25$ TeV, $S=0.3$ and $\tilde{g}=7$. This theory is roundly excluded simply by $S$ and the absence of a light higgs candidate but provides some reference values to place on the exclusion plot Figure~\ref{fig:collider}. It is not excluded purely in terms of the $\rho,A$ bounds.   

The minimal QCD scale up is already ruled out but we entertain here the possibility that a related theory with additional techniquark electroweak singlets can change the running so that the constraints on $S$ and the higgs mass can be accommodated. In terms of the runnings of $\alpha_{TC}$ in Figure~\ref{fig:running} for the $N_c=3$, $N_f=2$ case we would need to move the running from the left most profile (the two loop running for the theory) to the rather bizarre running shown one to the right. The one loop coupling scale has moved close to 700 TeV then (here by ``magic'' since we know this doesn't happen in QCD!) the IR is modified to create a very conformal IR fixed point that allows a light higgs. We also vary our parameter $\kappa$ to ensure $S=0.1$. As an example of the effects of these changes consider the $N_c=3, N_f=2$ theory with the matching to the new IR running performed when $\alpha=0.7$ - we find $M_A = 4.11$ TeV, $M_\rho = 3.63$  TeV, $F_A = 1.54$  TeV, and $F_\rho = 1.48$ TeV ($\omega = 0.047$ and is small as previously discussed - there is very little impact on the excluded regions from variation of this small size so we suppress discussion of it). 
{This spectrum is shown in the $\tilde{g}-M_A$ plane in Figure~\ref{fig:collider} together with the current LHC constraints - the mutated SU(3) point corresponds to the top red point.} In this mutation of QCD with exotic running the holographic model has predicted that the vector meson masses grow even further from exclusion by the LHC constraints. It seems reasonable that in a theory with strong coupling out to such a large scale the masses of the theory should be dragged to higher scales also. 

{We now perform this same analysis for varying $N_c$ and $N_f$ theories - it is possible that for one of these theories the IR running we impose is less fanciful. The spectrum comes from the  predictions of our holographic model tuned at each $N_c,N_f$ to give $S$=0.1 and the correct higgs mass.  The red colour is for $N_c=3$, green $N_c=4$ and blue $N_c=5$. The top edge of the box in each case corresponds to the one electroweak doublet theory result with the width representing an estimate of the theoretical error (we match the IR running at different values as described above in Section II). 

It is simple to also include the effects of additional electroweak singlets on top of a single doublet since they only affect the running of the UV coupling.  The points in Figure~\ref{fig:collider}  correspond to the motion of the right hand point on the top line (the one doublet result) for each colour (value of $N_c$) as the number of electroweak singlets is also changed to vary the UV running - the effect is small because the theories share much the same IR running to generate $m_h$.

In Figure~\ref{fig:collider} we have also extended the spirit of this analysis to theories with additional techniquark electroweak doublets that change the UV running in a known fashion (making the coupling run more slowly) and then adjusting the IR and $\kappa$ to match $S=0.1$ and the higgs mass. The extra doublets tend to increase $F_\Pi$ by $\sqrt{N_f}$ which reduces the overall mass scale. However, the need to reduce $S$ (which grows as $N_f$) leads to a tuning of $\kappa$ that increases the mass scale. The net result we find is that the mass scale of the mesons is largely unchanged. The decay constants $F_A$ though do scale as $\sqrt{N_f}$ so $\tilde{g}$ falls with the addition of further doublets. The results are again shown in Figure~\ref{fig:collider} - here one should move down in the coloured box associated with each $N_c$. Moving down the box corresponds to increasing the number of electroweak techni-doublets from one to $2 N_c$ where the theories are assumed to enter the conformal window. Note here the collider data is not directly applicable since it was generated for a single doublet theory but the masses of the mesons do appear beyond LHC also at this time.  

It is notable that the holographic models all lie on or near the line where $\mathcal{a}$, from the phenomenological model, is zero. The reason is that $\mathcal{a}$ parametrizes $\rho-A$ degeneracy and the points lie near the line $\mathcal{a}=0$ as a result of forcing a small $S$ parameter. In the holographic model where $\kappa$ is the only available parameter to tune this appears the unique solution. This makes it clear that in the phenomenological model much of the parameter space achieves $S$=0.1 by a complicated tuning of the two vector masses and their decay constants - it is not clear if these tunings are achievable in a UV complete model.

The broad conclusion of all of this analysis is that WTC models (if they exist) probably still lie well beyond the LHC's reach and are not yet fully excluded. 
At this point one has to again query how believable the running functions we have adopted are. Certainly the two loop runnings do include strongly coupled IR fixed points yet we should be sceptical of the fixed point values computed in this (gauge dependent) way. The spirit of the analysis, guessing an IR fixed point behaviour, is therefore not unreasonable but our runnings are hugely fine tuned (at one part in 100 in $N_f^{IR}$ which takes the value 11.43) to give the observed higgs mass. One might very reasonably conclude the chance of the real running falling on these tuned guesses is very low. On the other hand if such a tuned theory is the answer nature has chosen then one would encounter a light higgs and be able to deduce this tuning in the runnings! Here we do not wish  to advocate this latter view particularly but our results do show the bizarre nature of a technicolor theory that survives the current constraints and that more work is needed experimentally to exclude them completely.}

\begin{figure}[htb]
   \includegraphics[width=0.5\textwidth]{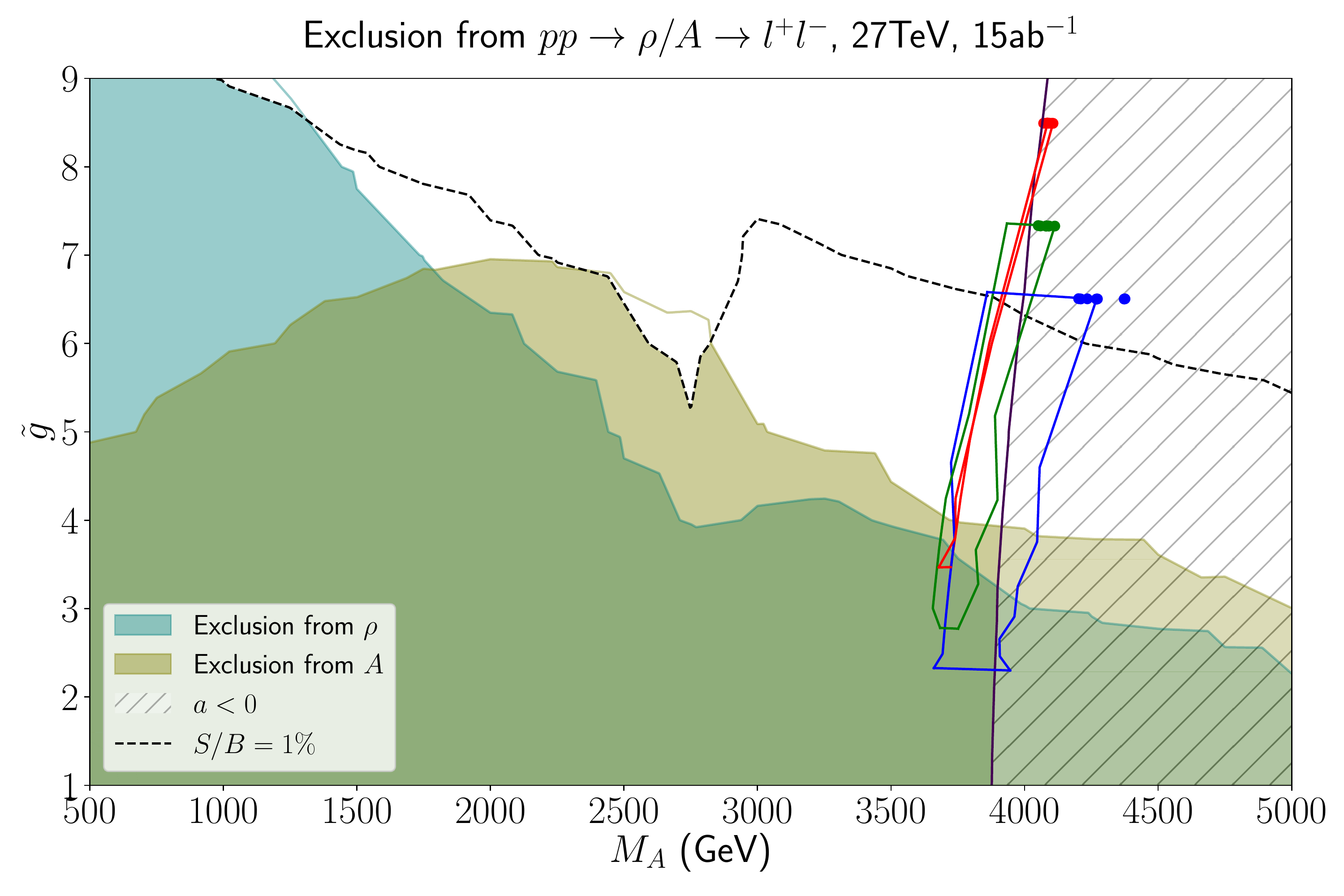}\\
   \includegraphics[width=0.5\textwidth]{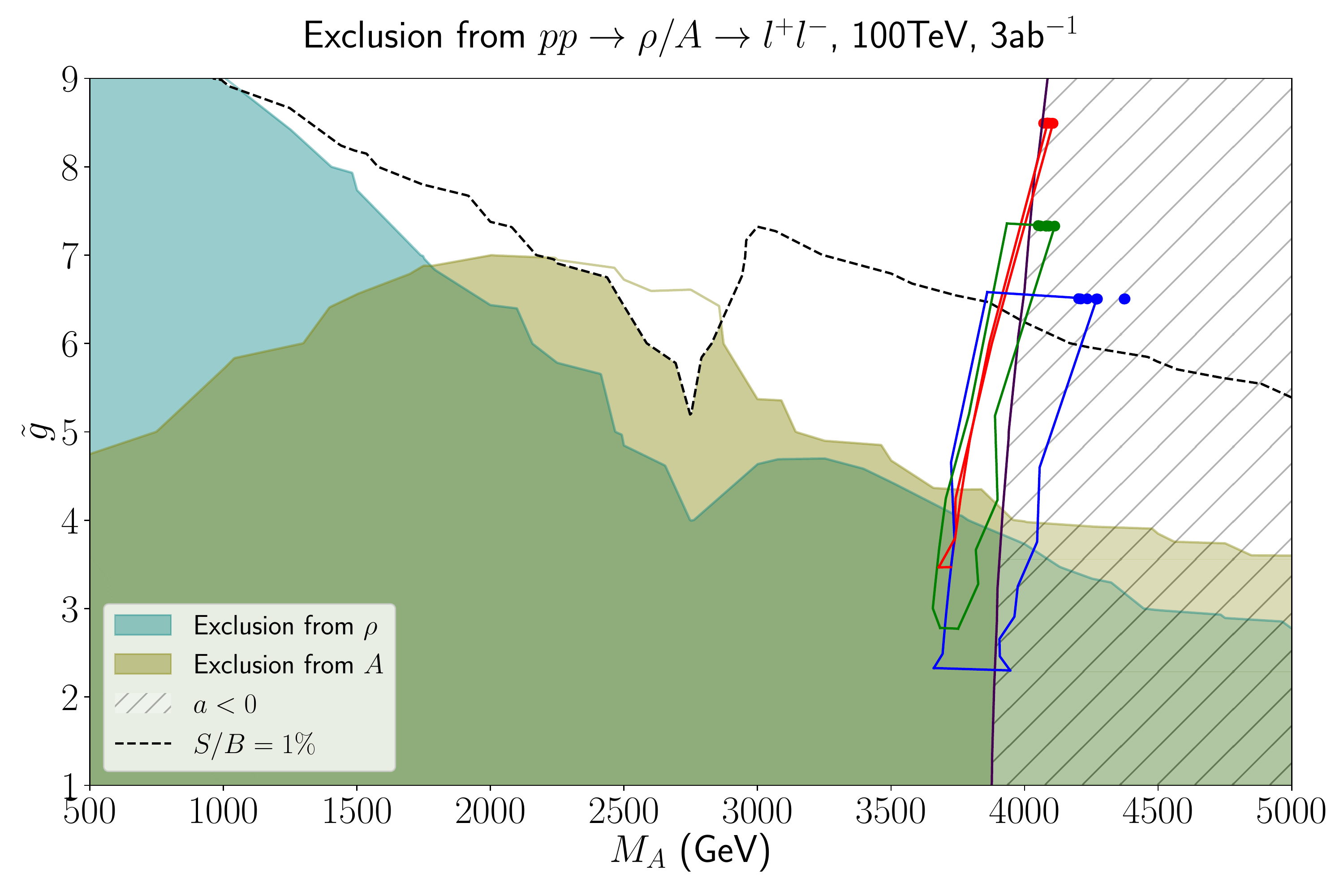}
  \caption{\label{fig:collider2}
  Shaded areas present 95\% CL projected exclusion  on the $M_A- \tilde{g}$ plane for  27(15 ab$^{-1}$) (top) and 100 TeV (3 ab$^{-1}$)(bottom) $pp$ collider from  dilepton DY resonance searches. The notations are the same as in Figure~\ref{fig:collider}.
  } \end{figure}
  
  \section{Beyond LHC}

{We have demonstrated that the LHC dilepton searches}  to date has not been able to exclude the WTC paradigm.  A total exclusion would need not only a higher collider energy but also new signatures to probe 4-5 TeV resonances especially in the large $\tilde{g}$ region
 with very low dilepton rates. We illustrate this point   in  Figure~\ref{fig:collider2} where we present projections for dilepton searches at  27(15 ab$^{-1}$) and 100 TeV (3 ab$^{-1}$) $pp$ collider. One can see a dramatic improvement of the sensitivity
 (in comparison to LHC@13TeV and LHC@14TeV\footnote{LHC@14 TeV with 3 ab$^{-1}$ would be able to reach  $\tilde g \simeq 4$ for $M_A$ around 4 TeV as demonstrated in~\cite{Belyaev:2018qye}.}) to the WTC parameter space at these future colliders, especially at  100~TeV where there is sensitivity to $\tilde g \simeq 4$ for $M_A$ around 4 TeV with a dilepton search. At the same time one can see that these searches would cover models only with large number of techni-doublets, while models coupling with $\tilde g \simeq 8$ are still far from reach even  at a 100~TeV collider if only the dilepton DY signature is used. 
 
 One can see that dilepton signature becomes less efficient in probing the WTC parameter space for large values of $\tilde g $ where the couplings of the $\rho/A$  to fermions are suppressed.
 Therefore exploration of higher values of $\tilde g $ motivates study of additional di-boson signatures
 either from  DY production or from the additional vector boson fusion (VBF) production channel.
 One should note that VBF production of $\rho/A$ followed by respective diboson(VV) or boson-higgs(VH) decay
 looks particularly promising in the very large  $\tilde g \simeq 8-9$ region since neither production nor decay of new heavy resonances are suppressed by $1/\tilde g$. 
 Moreover, the increase of collider energy can further enhance the significance of the VBF channel. An exploration of these additional $VV/VH$ signatures and VBF production channel, which could potentially cover the whole WTC parameter space,
 will be the subject of a follow-up paper.

 \section{Discussion}
  
The technicolor paradigm has long been appealing but it has been under fire for years from precision electroweak data and the discovery of a light higgs. Here we have asked the question of whether it can be finally put to bed by LHC data for searches for techni-$\rho$ and $A$ states. To declare a theory dead one must take the most conservative approach so we have entertained the idea that tuning the IR running of the theory may generate a sufficiently light higgs  (since we do not know which theory  might have such IR running we have imposed it on a range of theories with different $N_c, N_f$ in the hope to capture the true theory if it exists).  Holography provides a very simple analysis that predicts the techni-$\rho$ and $A$ spectrum and couplings based on the input running of $\gamma$,  the anomalous dimension of $\bar{q}q$,  and therefore provides a good first estimate of the mass spectrum of these theories. We have found that tuning $S$ to a small value naturally places these models on the $\mathcal{a}=0$ line of the phenomenological model that has been used previously for analysis. {Our main result shows that these models still lie beyond the reach of the LHC via Drell-Yan dilepton resonance searches. We have also shown that the DY signal alone will not exclude the most minimal models even at a 100~TeV (3 ab$^{-1}$) $pp$ collider which motivates future work on bringing additional signatures and production channels  at higher energy colliders that will exclude the paradigm.} 

The WTC models that survive here are fairly baroque, entering strong coupling at the 700~TeV or so scale and then running very slowly in the IR with the result that the resonances' masses are pushed up in scale. They also possess a large change to $\gamma=2$ near the fixed point so display the walking mechanism that pushes away the flavour scale. The biggest lesson perhaps to learn from this analysis is the difficulties that a light higgs leave for all Beyond the Standard Model theories  which now must possess IR fine tuning and and push new physics to high scales. On the other hand these models provide some motivation to build higher energy colliders and explore new signatures to fully probe the model parameter space.

\acknowledgements{
The authors are very grateful to
Emmanuel Olaiya for numerous useful discussions 
about the signal  fitting procedure and establishing the LHC limits.
{AB's and NE's work was supported by the
STFC  consolidated  grant  ST/P000711/1}.
AB acknowledges partial  support from Soton-FAPESP grant. AB also thanks the NExT Institute and   Royal Society International Exchange grant IE150682. AC acknowledges partial support from
SEPnet under the GRADnet Scholarship award.
}

\end{document}